\documentclass{kapproc}
\usepackage{graphicx}
\upperandlowercase
\kluwerbib
\begin{document}
\articletitle[]{Vertical distribution of stars and gas in a
galactic disk}
\author{Chanda J. Jog\\}
\affil{Indian Institute of Science, Bangalore 560012, India}
\email{cjjog@physics.iisc.ernet.in}
\begin{abstract}
We study the vertical density distribution of stars and gas (HI and H$_2$) 
in a galactic disk which is embedded in a dark mater halo. The new feature 
of this work is the inclusion of gas, and the gravitational coupling between 
stars and gas, which has led to a more realistic treatment of a 
multi-component galactic disk. The gas gravity is shown to be crucially 
important despite the low gas mass fraction. This approach physically 
explains the observed scaleheight distribution of all the three disk 
components, including the long-standing puzzle (Oort 1962) of a 
constant HI scaleheight observed in the inner Galaxy.
The above model is applied to two external galaxies: NGC 891 and NGC 4565, 
and the stellar disk is shown to be not strictly flat as was long believed but 
rather it shows a moderate flaring of a factor of $\sim 2$
within the optical radius. 
\end{abstract}


\section{Introduction}
The visible mass in a galactic disk is distributed in a thin
disk, hence the planar and the vertical dynamics can be treated
separately. The vertical distribution is not studied well unlike the 
planar mass distribution, though 
it is important to understand it
since it acts as a tracer of the galactic potential and the disk 
evolution.
The vertical distribution in a one-component, isothermal, self-gravitating disk was studied
in a classic paper by Spitzer (1942).
 However, the physical
origin of the thickness of stars and gas in a galactic disk and
in particular their radial variation is not yet fully understood- we 
address this problem here.

In the past, gas gravity was often ignored even when studying the
vertical distribution of gas, and the gas response to the
stellar potential alone was considered.
We show that stars and gas need to be treated jointly to get their
correct vertical distribution. Due to its
low velocity dispersion, gas forms a thin layer and hence
affects the dynamics and the vertical distribution of stars as
well, despite the low gas mass fraction.

\section {Formulation of Three-component Model}

The vertical scaleheight of HI gas in the inner Galaxy is
remarkably constant which has been a long-standing puzzle (Oort
1962, Dickey \& Lockman 1990). 
The vertical scaleheight is a measure of equilibrium
between the gravitational force and the pressure. Hence the HI gas responding to the
gravitational potential of an exponential stellar disk alone should show flaring, in contrast to the observed constant scaleheight.

We propose that an increase in the gravitational
force due to the inclusion of gas can decrease the gas scaleheight,
and study this idea quantitatively (Narayan \& Jog 2002b). We
formulate a general treatment where the stars, the HI gas and the H$_2$ gas
are treated as three gravitationally coupled disk components,
embedded in a dark matter halo. Thus each disk component
experiences the same joint potential due to the three disk
components and the dark matter halo, but each has a different
response due to its different random velocity dispersion. 

The equation of hydrostatic equilibrium along $z$ or the vertical direction for each
component and
the joint Poisson equation are solved together. Thus each
component obeys:
$${\frac {d^2 \rho_i}{d z^2}} \: = \: \frac {\rho_i}{<{(v_z)_i}^2>} \:
 \large[ - 4 \pi G (\rho_s + \rho_{HI} + \rho_{H_2}) + {\frac {d
(K_z)_{DM}}{d z}} \large] + {\frac {1}{\rho_i}} \large({\frac {d \rho_i} {d z} }\large)^2 $$
\noindent where $i$=1,2,3 denote the three disk components, and $\rho_i$ and
$<{(v_z)_i}^2>^{1/2}$ denote the mass density and  the velocity dispersion respectively.

The above three coupled, second-order differential equations
are solved numerically and iteratively, using $d \rho_i /d z =
0$ at the midplane and the observed mass surface density ($\Sigma_i$) as the
two boundary conditions in each case. This results in a 
sech$^2$-like density profile for the mass density in each case, and we use
its HWHM to define the vertical scaleheight.

\subsection {Resulting Vertical Scaleheight vs. Radius }
On using the joint potential, the HI scaleheight decreases as
expected, and is in a
better agreement with observations (Fig. 1),
especially at large radii where the effect of gas gravity
is higher because of the higher gas fraction.
\begin{figure}
\begin{center}
{\resizebox{6.5cm}{6.5cm}{\includegraphics[width=0.7\textwidth]{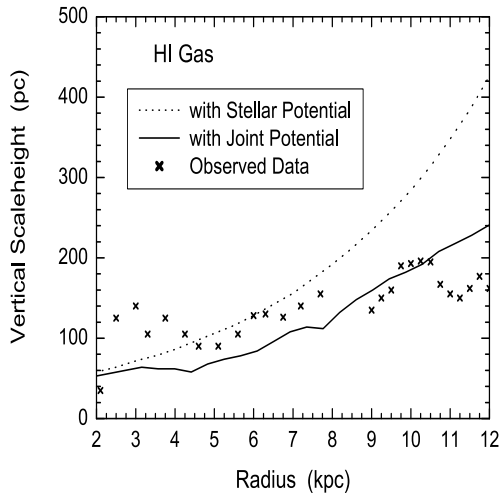}}}
\end{center}
\begin{center}
\caption{The joint potential gives a better fit to the HI data,
and explains the
constant HI scaleheight observed in the inner Galaxy. The agreement is even better for a slight linear increase in the HI velocity dispersion at small radii.}
\end{center}
\end{figure}
For H$_2$ gas, the joint potential gives  a good agreement at all radii.  
In the stellar case, the joint potential gives a nearly flat curve that 
slowly rises with radius with a slope of 24 pc kpc$^{-1}$ for 5-10 kpc, this 
agrees well with the near-IR observations (Kent et al. 1991).
\section{Radially Increasing Stellar Scaleheight}
We next applied this general model to two external, prototypical 
edge-on spiral galaxies: NGC 891 and NGC 4565, for which
gas density values are known from observations - see Narayan \& Jog (2002a) for details.

From the pioneering study of the vertical luminosity distribution
of edge-on spiral galaxies, van der Kruit \& Searle (1981a,b)
concluded that the stellar vertical scaleheight in a galactic disk is
constant with radius, and this is explained if the ratio $R_{vel} / R_d = 2$ where 
$R_{vel}$ and $R_d$ are the scalelengths with which the stellar velocity 
dispersion and disk luminosity fall off exponentially.
 This has been a paradigm in galactic
structure for 25 years. However, there is no clear physical
explanation for this. We argue that such a strict relation is 
 unrealistic in general given the different sources of
stellar heating now known to operate in a disk, such as
heating via
scattering off clouds or spiral arms within the galactic disk, and via 
heating  due to tidal encounters with other galaxies.

Hence, our model is applied to NGC 891 and NGC 4565, treating 
the above ratio as a free parameter. The
resulting stellar scaleheight is not strictly constant but instead shows 
flaring by a factor of $\sim 2$ within the optical radius (Fig. 2). This result agrees with observations (de Grijs \& Peletier 1997). This is an important result for galactic structure.
\begin{figure}
\begin{center}
{\resizebox{6.5cm}{6.5cm}{\includegraphics[width=0.7\textwidth]{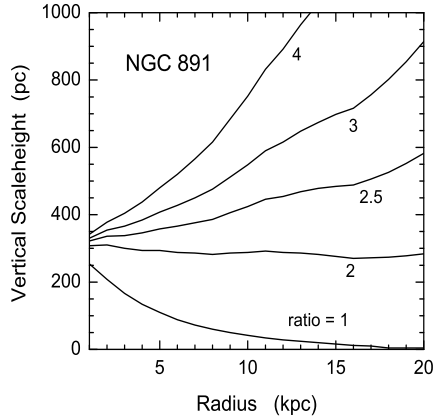}}}
\end{center}
\begin{center}
\caption[width=0.7\textwidth]{Resulting stellar vertical scaleheight vs. Radius: In general, the stellar scaleheight shows moderate flaring
when the ratio  $R_{vel} / R_d > 2$.}
\end{center}
\end{figure}
\section {Conclusions}
The multi-component
approach developed here cohesively explains the observed
vertical scaleheights of all the three disk components, namely
the stars, the HI and the H$_2$ gas in the inner Galaxy.

The above model is applied to two external galaxies, NGC 891 
and NGC 4565, and the resulting stellar scaleheight is shown to be not
constant with radius, instead it shows flaring. Such a moderate
increase in stellar scaleheight is likely to be common in other galaxies as well. 

\subsection{References}
\noindent Dickey, J.M., \& Lockman, F.J. 1990, ARA\&A, 28, 215

\noindent de Grijs, R., \& Peletier, R.F. 1997, A \& A, 320, L21

\noindent Kent, S.M., Dame, T.M., \& Fazio, G. 1991, ApJ,  378, 131

\noindent  Narayan, C.A., \& Jog, C.J. 2002 a, A \& A, 390, L35             

\noindent Narayan, C.A., \& Jog, C.J. 2002 b, A \& A, 394, 89

\noindent Oort, J.H. 1962, in The Distribution and Motion of Interstellar Matter in Galaxies, IAU symp. 15, ed. L. Woltjer (New York: Benjamin), 3

\noindent Spitzer, L. 1942, ApJ, 95, 329

\noindent van der Kruit, P.C., \& Searle, L. 1981 a,b , A \& A, 95, 105 and 116


\end{document}